\documentclass[structabstract]{aa} 

\usepackage{graphicx}
\usepackage{amsmath}
\usepackage{txfonts}
\usepackage{natbib}

\makeatletter
\newlength{\figwidth}
\if@referee
\else
\usepackage{color}

\fi
\makeatother

\newcommand{\abs}[1]{\ensuremath{\left\lvert #1\right\rvert}}

\newcommand{\be}{\begin{equation}}
\newcommand{\ee}{\end{equation}}
\newcommand{\bs}{\begin{subequations}}
\newcommand{\es}{\end{subequations}}

\newcommand{\pa}{\ensuremath{_\parallel}}
\newcommand{\se}{\ensuremath{_\perp}}

\newcommand{\De}{\ensuremath{\varDelta}}
\newcommand{\Om}{\ensuremath{\varOmega}}

\newcommand{\Ga}{\ensuremath{\varGamma}}
\newcommand{\Si}{\ensuremath{\varSigma}}

\newcommand{\f}[1]{\ensuremath{\boldsymbol{#1}}}
\newcommand{\m}[1]{\ensuremath{\left\langle #1\right\rangle}}

\newcommand{\df}{\ensuremath{\mathrm{d}}}

\begin{document}
\title{On the ergodicity of perpendicular cosmic ray transport}
\author{R.\,C. Tautz}
\institute{Zentrum f\"ur Astronomie und Astrophysik, Technische Universit\"at Berlin, Hardenbergstra\ss{}e 36, D-10623 Berlin, Germany\\\email{robert.c.tautz@gmail.com}}

\date{Received February 26, 2016; accepted April 30, 2016}

\abstract
{}
{The random walk of energetic charged particles in turbulent magnetic fields is investigated. Special focus is placed on transport across the mean magnetic field, which had been found to be subdiffusive on many occasions. Therefore, a characterization using the concept of ergodicity is attempted by noting the connection to the time evolution of the mean-square displacement.}
{Based on the test-particle approach, a numerical Monte-Carlo simulation code is used to integrate the equation of motion for particles that are scattered by magnetic turbulence. The turbulent fields are generated by superposing plane waves with a Kolmogorov-type power spectrum. The individual particle trajectories are then used to calculate a variety of statistical quantities.}
{The simulation results clearly demonstrate how the heterogeneity of the particle ensemble causes the system to be weakly non-ergodic. In addition, it is shown how the step length distribution varies with the particle energy. In conclusion, cross-field transport is non-Gaussian but still almost diffusive.}
{}

\keywords{Diffusion --- Turbulence --- Magnetic fields --- (ISM:) cosmic rays --- Methods: numerical}
\authorrunning{Tautz}
\titlerunning{Ergodicity of perpendicular transport}
\maketitle

\section{Introduction}

Describing the stochastic motion of energetic charged particles due to the interactions with turbulent (electro-)magnetic fields has been a long-standing problem in the fields of high-energy astrophysics and (laboratory) plasma physics \citep[see, e.\,g.,][for an introduction]{rs:rays,sha09:nli}. A prominent example is that of cosmic rays being scattered in the fluctuating magnetic fields of the Milky Way and the heliosphere, depending on their kinetic energy. On smaller scales, particles interacting with planetary magnetospheres and the prediction of space weather are important, not least for the safety of electronic devices \citep[e.\,g,.][]{sch05:wea,bot06:wea}.

By neglecting binary Coulomb collisions, a Fokker-Planck and subsequently a diffusion-convection equation can be derived from the Vlasov equation. The diffusion coefficients are then solely determined by the properties of the magnetic fields. Considerably effort has been put into the analytical derivation of the diffusion coefficients, and numerous approaches---both analytical and numerical---have been invoked \citep[e.\,g.,][]{sha09:nli,tau13:num}. An important parameter for the classification of transport processes \citep{bak15:sto} is the Kubo number \citep{kub63:lio,zim00:kub,gus11:clu} which, for fluid turbulence, is defined as the relative turbulence strength multiplied with the ratio of the correlation lengths along and across the mean magnetic field $R=(\delta B/B_0)(\ell\pa/\ell\se)$.

In recent years, increasing evidence has been found that anomalous transport may play an important role in astrophysics \citep[e.\,g.,][]{zim06:dif,per08:sup,kla08:ano,spa08:sto,tau10:sub}. Accordingly, both a theoretical description of the underlying physics as well as a detailed characterization of the data---both numerically and observationally obtained---is required. A general proof that the deflections induced by turbulent magnetic fields indeed cause a diffusive behavior is still elusive. This problem is of interest both for theoretical investigations as well as for the analysis of data for instance taken in situ by spacecraft. A non-diffusive behavior results in time-dependent (``running'') diffusion coefficients, in which case the solution of the diffusion equation becomes considerably more involved and may even be replaced by a fractional differential equation \citep[e.\,g.,][and references therein]{met04:ano,tau16:pro}, In addition, non-ergodic behavior emphasizes the individuality of the particles,
 thus requiring more care when conclusions are drawn based on a limited data set.

Here, the transport of energetic charged particles---in particular, cosmic rays---will be investigated based on the trajectories of individual particles \citep[see also][]{met12:tra}. Special focus is placed on cross-field transport, the importance of which had often been overlooked. Recently, the role of subdiffusive perpendicular diffusion (e.\,g., \citealt{tau10:sub}, but cf. \citealt{qin02a:rec,xus13:tra}) was emphasized in the extraction of diffusion coefficients from intensity profiles \citep{tau16:pro}. Generally, the necessity of anisotropic diffusion is increasingly recognized in the community \citep[e.\,g.,][]{kis14:pcd,gir14:ani}.

This article is organized as follows. In Sec.~\ref{tra}, the mean-square displacement and its importance in the context of random walks and turbulent transport is introduced. The numerical simulations that are used to calculate the transport of charged energetic particles in turbulent magnetic fields are briefly described in Sec.~\ref{sim}. The results presented in Sec.~\ref{msd} comprise several quantitative criteria and distributions that are based on the mean-square displacement. In Sec.~\ref{heter}, results are shown for the ensemble heterogeneity, which is used to characterize the typical behavior of random walks. Sec.~\ref{summ} provides a short conclusion.

\section{Turbulent transport processes}\label{tra}

The mathematical framework developed to treat the random walk \citep[see, e.\,g.,][for an introduction]{cha43:sto} of an ensemble of identical particles is based on the similarity to the diffusive mixing of gases and liquids \citep{fic55:dif}. For further applications such as the analytical or numerical treatment of shock acceleration, diffusivity is then often simply assumed. 

Due to the vanishing mean of the particle's displacement, one readily requires the second moment, which leads to the mean-square displacement (MSD). Based on the concept of ergodicity \citep{gus11:clu}, which states that ensemble averages and time averages should be equal, two versions are widely used. The time-averaged MSD is defined as
\be\label{eq:msqT}
\overline{\De^2(t)}=\frac{1}{T-t}\int_0^{T-t}\df t'\left[x(t'+t)-x(t')\right]^2
\ee
for each particle individually. Note that the ``running'' time coordinate during the process is denoted as $t$ in order to distinguish it from the total simulation time or time span of the measurement, $T=\max(t)$. In the simulations presented in Sec.~\ref{msd}, the measurement time $T$ is always chosen such that parallel transport---with respect to the mean magnetic field---has already become diffusive. This is important since strictly speaking diffusion applies only in the infinitely long time limit.

For ergodic processes, the time-averaged MSD for individual particles is equal to the ensemble-averaged MSD as
\be\label{eq:ergodicity}
\overline{\De^2(t)}=\m{\De x(t)^2}\sim t^{\,\alpha}
\ee
which both are proportional to a power law in $t$. If the two functions are not equal, the process is generally characterized as \emph{weakly non-ergodic} to distinguish it from strongly non-ergodic processes, in which case the phase space of the particles is separated.

As will be shown in Sec.~\ref{heter}, the time-averaged MSD typically has a large variance for (weakly) non-ergodic processes, which reflects the heterogeneity of the individual particles. Therefore, another variable can be introduced, which is the ensemble-averaged, time-averaged MSD defined through
\be\label{eq:msqTE}
\m{\overline{\De^2(t)}}=\frac{1}{N}\sum_{i=1}^N\overline{\De_i^2(t)}.
\ee

To describe the transport of energetic charged particles and to evaluate the time evolution of the probability distribution function, the diffusion coefficients are calculated via $\kappa=\langle\De x(t)^2\rangle/(2t)$ and are inserted in a transport equation such as the \citet{par65:pas} or \citet{roe69:int} equation. In such cases a diffusive behavior---i.\,e., $\alpha=1$ in Eq.~\eqref{eq:ergodicity}---is implicitly assumed. There are, however, indications that anomalous diffusion may play an important role in many scenarios, including the propagation of solar energetic particles toward Earth \citep[e.\,g.,][]{abl16:sol}.

Much effort has been put into characterizing the process in terms of a waiting time distribution \citep[e.\,g.,][and references therein]{mer15:sub}. Originally, a continuously distributed waiting time was used to describe the discrete motion charge carriers in amorphous semiconductors. However, the concept is difficult to apply for the turbulent transport of charged particles, which---at least in the absence of electric fields---move at a constant speed. Instead, the distribution of step lengths may be better suited as shown in Sec.~\ref{msd:steps}.

\section{Monte-Carlo simulation}\label{sim}

Measurements in the solar winds \citep[see, e.\,g.,][for an overview]{bru05:sol} have revealed that the Fourier spectrum of the turbulent fields is roughly in agreement with Kolmogorov's prediction \citep{kol41:tur}. The turbulent magnetic fields therefore can be best modeled in Fourier space using a kappa-type power spectrum \citep{sha09:flr} as
\be\label{eq:spect}
G(k)\propto\frac{\abs{\ell_0k}^q}{\left[1+(\ell_0k)^2\right]^{(s+q)/2}},
\ee
where $q=3$ is the energy range spectral index \citep[cf.][]{gia99:sim} for isotropic turbulence. The turbulence bend-over scale, $\ell_0\approx0.03$\,au, reflects the transition from the energy range $G(k)\propto k^q$ to the Kolmogorov-type inertial range, where $G(k)\propto k^{-s}$ with $s=5/3$.

In the following, the numerical Monte-Carlo code \textsc{Padian} will be used to evaluate the properties of perpendicular transport for the turbulence model described above. A general description of the code and the underlying numerical techniques can be found elsewhere \citep[see also \protect\citealt{mic96:alf,gia99:sim,lai13:cro}]{tau10:pad}. Specifically, the turbulent magnetic fields are generated via a superposition of $N$~plane waves as
\be\label{eq:dB}
\delta\f B(\f r,t)=\sum_{n=1}^N\f e'_\perp\sqrt{G(k_n)\,\De k_n}\,\cos\!\left(k_nz'+\beta_n\right),
\ee
where the wavenumbers $k_n$ are distributed logarithmically in the interval $k_{\text{min}}\leqslant k_n\leqslant k_{\text{max}}$ and where $\beta$ is a random phase angle. The polarization vector is chosen so that $\f e'_\perp\cdot\f e'_z=0$ with $\f k\parallel\f e'_z$, respectively, which ensures the solenoidality condition. For isotropic turbulence, the primed coordinates are determined for each $n$ by a rotation matrix with random angles.

From the integration of the equation of motion, the parallel diffusion coefficient, $\kappa\pa$, and the parallel mean-free path, $\lambda\pa$, can be calculated by averaging over an ensemble of particles and by determining the MSD in the direction parallel to the background magnetic field as $\kappa\pa=(v/3)\lambda\pa=\langle(\varDelta z)^2\rangle/(2t)$. For the perpendicular diffusion coefficient, a similar procedure performed but, in addition, the results are averaged over the $x$ and $y$ directions. It is important to note that realistic and robust results can be obtained only if a number of turbulence realizations (i.\,e., sets of random numbers) is considered, over which the MSD needs to be averaged. This fact will be revisited in Sec.~\ref{heter}.

The dynamics of charged particles that interact with magnetic fields are determined through their rigidity $pc/q$, i.\,e., the momentum per charge. Therefore, the \textsc{Padian} code uses a normalized rigidity variable $R=\gamma v/\Om\ell_0$, where $\gamma=(1+v^2/c^2)^{-1/2}$ is the relativistic Lorentz factor and $\Om=qB/mc$ is the gyrofrequency with $B$ the strength of the mean magnetic field. In addition, note that, throughout this paper, all lengths are normalized to the turbulence bend-over scale, i.\,e., $x=x_{\text{phys}}/\ell_0$ and all times are normalized to the gyrofrequency, i.\,e., $t=\Om t_{\text{phys}}$. Here, the index ``phys'' denotes physical lengths and times given in centimeters and seconds, respectively. Consequently, diffusion coefficients are given as $\kappa=\kappa_{\text{phys}}/(\ell_0^2\Om)$.

\begin{figure}[tb]
\centering
\includegraphics[width=\figwidth]{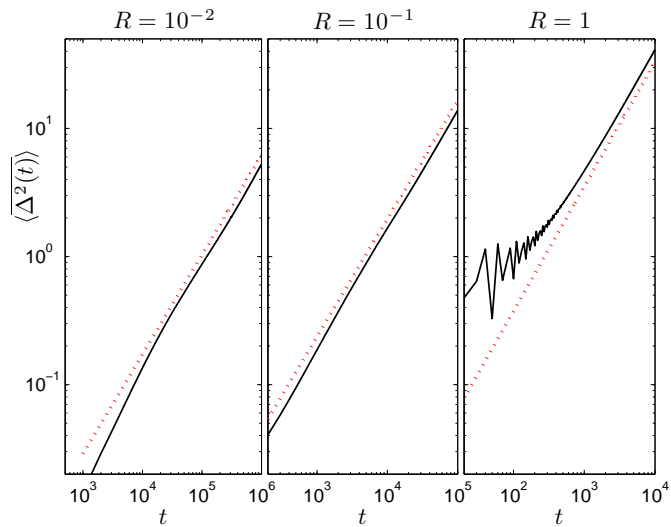}
\caption{Walk dimension as obtained from the time-averaged, ensemble averaged MSD defined in Eq.~\eqref{eq:msqTE}. The black solid lines show the simulation results, while the red dotted lines show the power law given in Eq.~\eqref{eq:dimw} with the best fit values as summarized in Table~\ref{ta:dim}.}
\label{ab:alphafit}
\end{figure}

The major importance of such simulations is found in the direct calculation of the transport parameters, as opposed to simulations that employ an---often simplified---model for the diffusion coefficients. In addition, the diffusion coefficients show a non-trivial time dependence, due to which at least three separate phases can be distinguished \citep[see, e.\,g.,][]{tau11:nlg,pro15:bal,abl16:sol}.

\section{Results for the MSD}\label{msd}

\begin{table*}[bth]
\centering
\begin{tabular}{llllll}\hline\hline
$R$						& $10^{-2}$					& $10^{-1}$					& $1$\\\hline
$d_{\text w}$				& $2.577\pm7.25\times10^{-3}$		& $2.173\pm9.44\times10^{-3}$ 	& $2.057\pm3.16\times10^{-4}$\\
$2/d_{\text w}$			& $0.776\pm2.19\times10^{-3}$		& $0.920\pm3.99\times10^{-3}$		& $0.97\pm4.57\times10^{-2}$\\
$d_{\text s}$				& $1.447\pm3.17\times10^{-4}$		& $1.843\pm4.55\times10^{-3}$		& $1.944\pm8.46\times10^{-4}$\\
$d_{\text s}/2$			& $0.7235\pm1.58\times10^{-4}$	& $0.9215\pm2.27\times10^{-3}$	& $0.972\pm4.23\times10^{-4}$\\\hline\hline
\end{tabular}
\caption{Confirmation of the Alexander-Orbach relation as obtained by connecting the respective walk and spectral dimensions according to Eq.~\eqref{eq:alex}. For the three rigidity values, the relevant values are shown together with their standard deviations\protect\footnotemark[1]. The fractal dimension is taken to be $d_{\text f}=2$ because only perpendicular transport is considered here.}
\label{ta:dim}
\end{table*}

\footnotetext[1]{Note that the errors solely reflect the fit uncertainties and thus neglect the additional degree of freedom arising from manually choosing the relevant time range.}
\stepcounter{footnote}

In the following two sections, the results from a Monte-Carlo simulation as described in Sec.~\ref{sim} will be presented. Unless stated otherwise, three representative values are assumed for the particle rigidity, which are $R=10^{-2}$, $10^{-1}$, and $1$.

\subsection{Alexander-Orbach relation}

For the random walk on clusters of fractals, \citet{ale82:fra} showed that there is a relation connecting three dimensions relevant for the random walk of particles in an arbitrary domain. A confirmation of this conjecture has proven to be challenging \citep[e.\,g.,][]{ley83:fra,nak03:fra}. For this reason, here its validity will be tested for an entirely different system than originally anticipated.

The relevant dimensions include first the fractal dimension $d_{\text f}$ which, in the present case, is simply the number of spatial dimensions, i.\,e., $d_{\text f}=2$ for perpendicular transport. Second, the walk dimension is obtained from the MSD via
\be\label{eq:dimw}
\m{\De x(t)^2}\sim t^{2/d_{\text w}}
\ee
so that $d_{\text w}=2$ would indicate a diffusive walk. The fit to the simulation results of the perpendicular MSD is shown in Fig.~\ref{ab:alphafit} with the fit parameters listed in Table~\ref{ta:dim}. To a good accuracy, the time evolution of the MSD can be described as a power law, at least after the initial transient effects (the so-called ballistic phase in which particles move freely until they are scattered for the first time). Note also that the time-averaged, ensemble-averaged MSD is identical with the usual definition of the MSD. For low rigidities, one notes that the particles have moved at most a few bend-over scales, which scale is proportional to the turbulence correlation length \citep[see, e.\,g.,][]{sha09:nli}. Formally, however, a diffusive description can be applied only in the limit of large times when the particle has traveled several bend-over scales so that its motion is no longer correlated with the initial conditions. Even though such a simulation is computationally very demanding, several tests were performed, which did not reveal any later variations in the slope of the MSD for larger $T$, thus suggesting that a shorter simulation time is sufficient.

Third, the spectral dimension $d_{\text s}$ is related to the probability density at the origin---i.\,e., the source---of perpendicular transport, which here is the $z$ axis. The relevant quantity to be determined from the simulated particle trajectories is thus
\be\label{eq:dims}
P(z=0,t)\sim t^{-d_{\text s}/2},
\ee
which can be obtained by recording the number (density) of particles in the form of an intensity profile\footnote{Note that, unlike in previous work \citep{tau16:pro,abl16:sol}, here no kernel function is used. The reasons are that: (i) near the $z$ axis, a sufficiently large number of particles will be registered at all times; (ii) potential inaccuracies due to the specific choice of a kernel will be avoided altogether.} \citep{tau16:pro}.

\begin{figure}[tb]
\centering
\includegraphics[width=\figwidth]{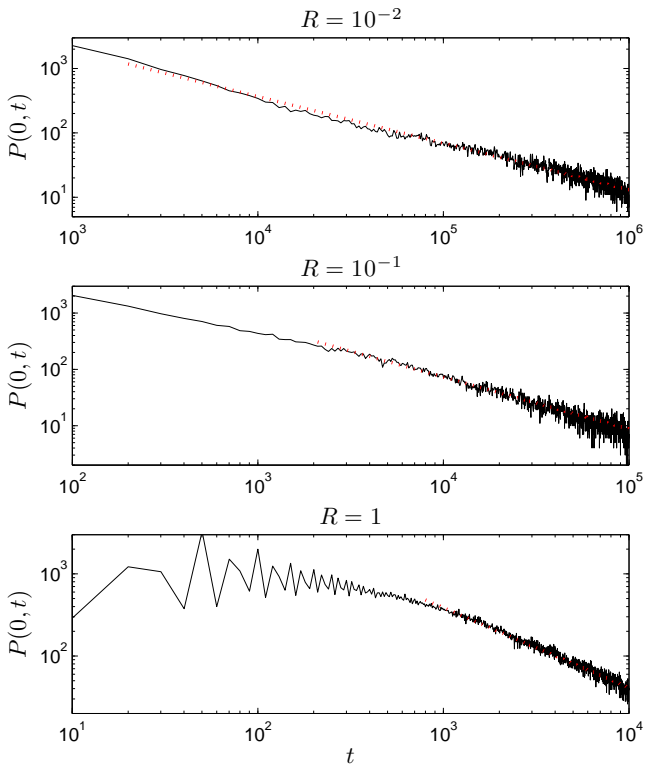}
\caption{Spectral dimension as obtained from the decreasing particle flux at the $z$ axis, which illustrates the turbulent transport in the perpendicular directions. For the three rigidities, the power law given in Eq.~\eqref{eq:dims} is best fitted the values summarized in Table~\ref{ta:dim}.}
\label{ab:profileOrigin}
\end{figure}

The result is shown in Fig.~\ref{ab:profileOrigin}. While for ergodic transport and from the diffusion equation a value of $d_{\text s}=2$ is expected, it can be seen that this value is not exhibited in particular for low rigidities. Instead, a reduced slope is observed, which is indicative of a subdiffusive process and reflects the fact that the particles' residence time near the origin is increased.

The Alexander-Orbach relation connects these three dimensions and states that
\be\label{eq:alex}
d_{\text s}/2=d_{\text f}/d_{\text w}=\alpha.
\ee
By comparing Figs.~\ref{ab:alphafit} and \ref{ab:profileOrigin}, it can be seen that the simulation results well confirm the validity of Eq.~\eqref{eq:alex}. Accordingly, two independent measurements confirm that $\alpha$ decreases with increasing rigidity (see Table~\ref{ta:dim}).

\begin{figure}[tb]
\centering
\includegraphics[width=\figwidth]{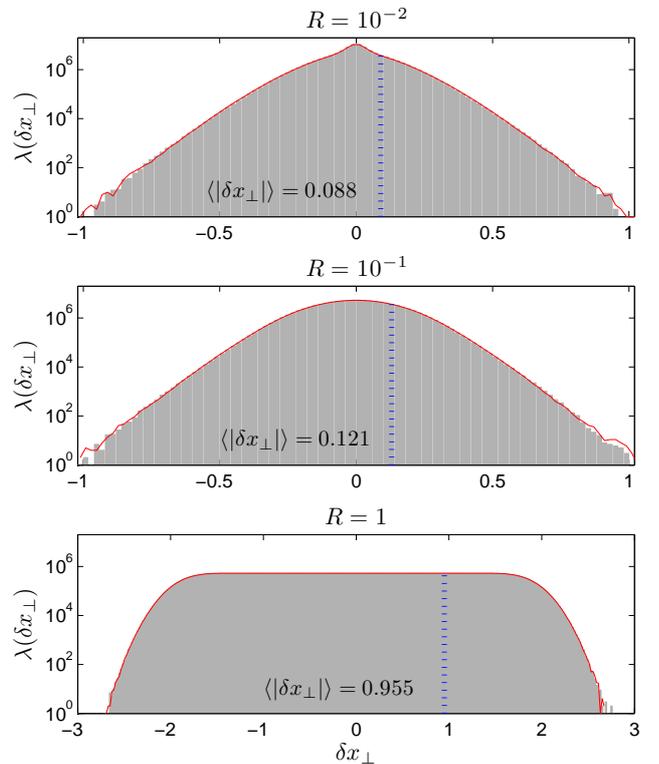}
\caption{Distribution of steps, $\psi(\De x\se)$, for a simulation with three rigidity values. The gray areas and the red lines illustrate the distribution of step lengths in $x$ and $y$ direction, respectively. In addition, the mean (positive) step length is depicted by the blue dotted lines with the values given directly in the figure.}
\label{ab:stepdistri}
\end{figure}

The comparison of the walk and spectral dimensions is important to judge the general behavior: if $d_{\text w}\geqslant d_{\text f}$---as is the case here---then the exploration of space is compact, which means that the particle will visit each point in space multiple times for an infinitely duration of the random walk. In the direction perpendicular to the mean magnetic field, therefore, particles tend to remain confined instead of diffusively filling space.

\subsection{Step lengths distribution}\label{msd:steps}

\begin{table*}[bth]
\centering
\begin{tabular}{llllll}\hline\hline
$R$											& $10^{-2}$							& $10^{-1}$							& $1$\\\hline
$\langle\delta x\se^2\rangle(T)$					& $1.5\times10^{-2}\pm2.04\times10^{-3}$	& $2.3\times10^{-2}\pm3.2\times10^{-3}$ 	& $1.3\pm0.17$\\
$\langle\delta x\se^2\rangle(T)/2\langle\tau\rangle$	& $7.3\times10^{-6}\pm1.02\times10^{-6}$	& $1.2\times10^{-4}\pm1.6\times10^{-5}$		& $6.1\times10^{-2}\pm8.3\times10^{-3}$\\
$\kappa\se(T)$									& $2.9\times10^{-6}\pm9.4\times10^{-7}$		& $6.9\times10^{-5}\pm1.43\times10^{-5}$	& $2.1\times10^{-3}\pm2.56\times10^{-4}$\\
$\max(\kappa\se)$								& $7.3\times10^{-6}\pm2.6\times10^{-6}$		& $1.1\times10^{-4}\pm2.36\times10^{-5}$	& $6.1\times10^{-2}\pm7.61\times10^{-3}$\\\hline\hline
\end{tabular}
\caption{MSD as derived from the step length distribution. The diffusion coefficient as obtained from the ratio of the MSD and the characteristic time step is in good agreement with the maximum of the time-dependent diffusion coefficient, which is directly calculated from the simulated particle trajectories.}
\label{ta:step}
\end{table*}

In addition to the waiting times, an equally important distribution is that of the step lengths, hereafter denoted as $\psi(\De x\se)$. Here, this distribution describes the probability for a certain displacement within the (constant) time step $\tau$. Note that this time step does not correspond to the (adaptively refined) time step used by the differential equation solver; instead, $\tau$ is the arbitrarily chosen spacing of the simulation output.

\begin{figure}[tb]
\centering
\includegraphics[width=\figwidth]{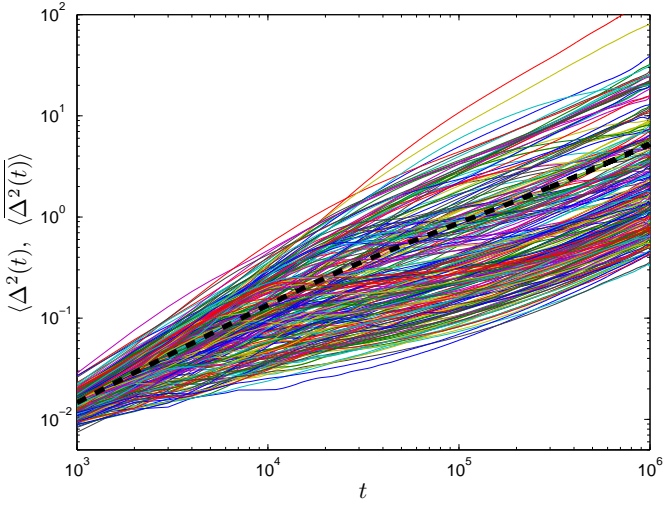}
\caption{Sample results for the time-averaged MSD as a function of the time lag, $t$, which is the time variable during the simulation. The dashed black line shows the ensemble-averaged, time-averaged MSD, which can be fitted to a power law $\propto t^{0.7235}$ for $t\gg t_{\text{ball}}$, where $t_{\text{ball}}$ is the ballistic time scale. The colored lines show the ensemble-averaged, time-averaged MSD for individual turbulence realizations.}
\label{ab:variety}
\end{figure}

Together with the mean (positive) step length, the step distributions are shown in Fig.~\ref{ab:stepdistri}. The variations in the results for different particle rigidities are caused by the turbulence power spectrum, which introduces an absolute scale---the bend-over scale, $\ell_0$---that has to be related to the particle Larmor radius, $R_{\text L}$. For higher rigidities, it is significantly less likely for a particle to become trapped in small-scale structures and, at the same time, the particle motion becomes increasingly random and unpredictable. These findings are reflected in the fact that the distribution $\psi$ is flat over a large range of step lengths, as illustrated in Fig.~\ref{ab:stepdistri}.

From the step length distribution, the diffusion constant can be inferred as \citep[e.\,g.,][]{kla87:ano}
\be\label{eq:kappa}
\kappa\se=\frac{\m{\delta x\se^2}}{2\m\tau},
\ee
where $\langle\tau\rangle$ would normally be the characteristic waiting time. For a continuously moving particle---particularly if its speed is constant---this quantity can be simply the time step at which the displacement is recorded. The results are summarized in Table~\ref{ta:step}, confirming the validity of Eq.~\eqref{eq:kappa} provided that the maximum of the time-dependent diffusion coefficient is taken into account.

\section{Ensemble heterogeneity}\label{heter}

\begin{figure}[tb]
\centering
\includegraphics[width=\figwidth]{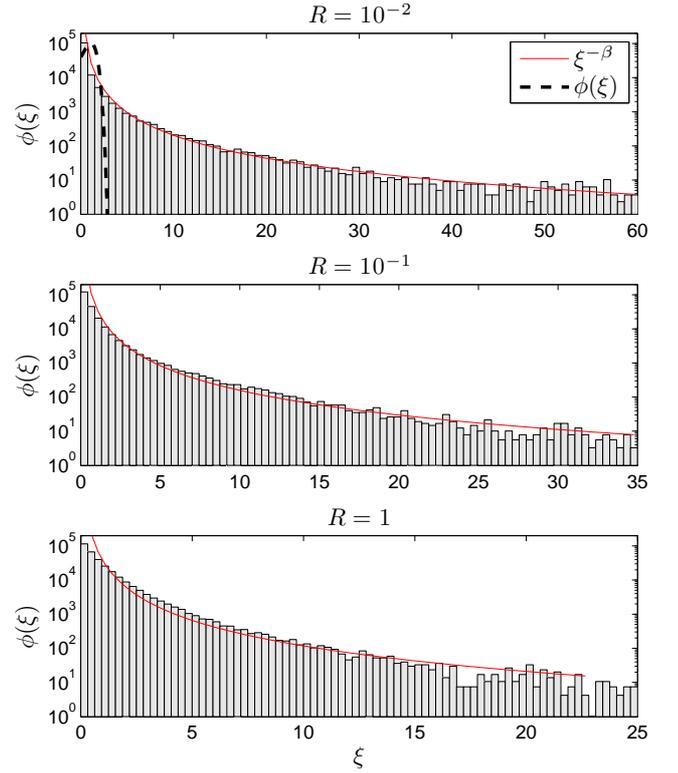}
\caption{Distribution of the MSD randomness variable $\xi$ as defined in Eq.~\eqref{eq:phi}. For the three rigidity values, the histograms of the numerical simulation are shown in logarithmic units. Note that there are significantly larger outliers, which are omitted here for clarity. The red solid lines show power law fits $\propto\xi^{-\beta}$ with $\beta=2.226$ (upper panel), $\beta=2.404$ (middle panel), and $\beta=2.486$ (lower panel). The black dashed line in the upper panel shows the distribution $\phi(\xi)$ as defined in Eq.~\eqref{eq:phi}.}
\label{ab:phi_distri}
\end{figure}

As stated in Sec.~\ref{tra}, ergodic behavior requires an ensemble of individual particles to move in a statistically independent manner in entirely random directions \citep{met14:erg}. Reversing the argument, ergodicity breaking is caused by particles with a time-average that deviates from that of the ensemble. In Fig.~\ref{ab:variety}, the time-averaged and ensemble-averaged MSDs are shown for charged particles being scattered by a stochastic magnetic field. Notably, the ensemble-averaged MSD agrees with the time-averaged MSD as defined in Eq.~\eqref{eq:msqT} only if the latter is also ensemble-averaged. Therefore, Eq.~\eqref{eq:ergodicity} is not fulfilled for perpendicular cosmic-ray transport.

\begin{figure}[tb]
\centering
\includegraphics[width=\figwidth]{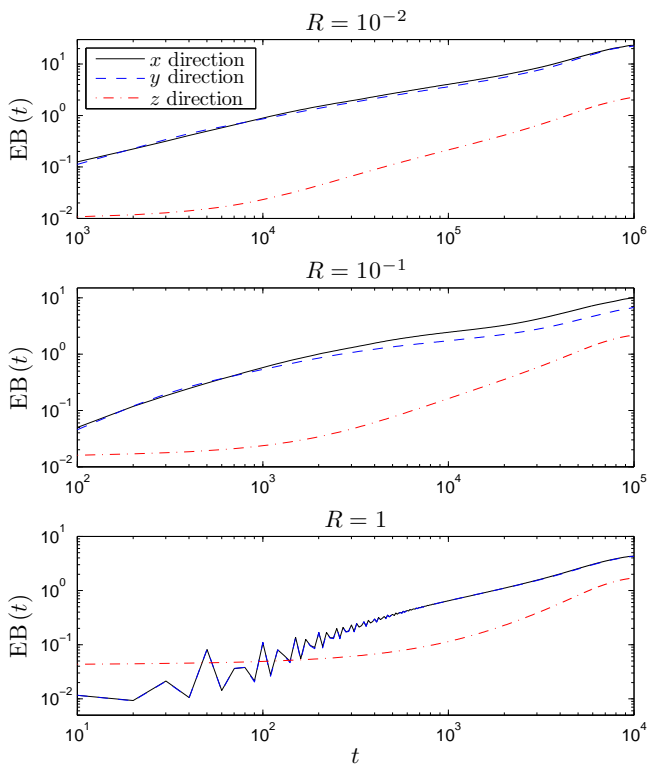}
\caption{Ergodicity breaking parameter as defined in Eq.~\eqref{eq:EB} for the three rigidity values. The black solid and blue dashed lines (with the latter being only marginally visible in the bottom panel) show $\text{EB}(t)$ for the $x$ and $y$ directions, respectively. For comparison purposes, the red dot-dashed line shows $\text{EB}(t)$ for the $z$ direction.}
\label{ab:break}
\end{figure}

Note that the various curves in Fig.~\ref{ab:variety} do not show the time-averaged MSD for individual particles. Instead, all curves represent averages over different sub-ensembles, each for a given turbulence realization. This result therefore illustrates the strong effect of a particular turbulence realization, which has been pointed out before by \citet{mer15:rea}.

\subsection{Heterogeneity distribution}

To account for the ensemble heterogeneity in a more quantitative way, a dimensionless variable can be introduced as
\be
\xi=\left.\overline{\De^2(t)}\right/\m{\overline{\De^2(t)}},
\ee
which is based on the MSDs defined in Eqs.~\eqref{eq:msqT} and \eqref{eq:msqTE}. Its probability distribution has been derived analytically \citep{hey08:msq} to be
\be\label{eq:phi}
\phi(\xi)=\frac{\Ga^{1/\alpha}(1+\alpha)}{\alpha\xi^{1+1/\alpha}}\,g_\alpha\left(\frac{\Ga^{1/\alpha}(1+\alpha)}{\xi^{1/\alpha}}\right),
\ee
where $\alpha$ is the slope of the MSD as given in Eq.~\eqref{eq:ergodicity}.

The function $g_\alpha(x)$ is defined through a Laplace transformation as
\be
\int_0^\infty\df x\;e^{-px}g_\alpha(x)=\exp\left(-p^\alpha\right)
\ee
for $p>0$ and $\alpha\in(0,1)$. For rational values $\alpha=\ell/k$ and $x\geqslant0$, the solution is \citep{pen10:lev}
\be\label{eq:levy}
g_{\ell/k}(x)=\frac{\sqrt{k\ell}}{x\left(2\pi\right)^{(k-\ell)/2}}\,G_{\ell,k}^{k,0}\left(\left.\frac{\ell^\ell}{k^kx^\ell}\right\rvert
\begin{matrix}
\Si(\ell,0)\\
\Si(k,0)
\end{matrix}
\right)
\ee
where the Meijer~$G$ function\footnote{Conveniently, the Meijer~$G$ function is already implemented in software packages such as Mathematica.} \citep{bea13:mei} takes as parameters two special lists of elements defined as $\Si(k,a)=a/k,(a+1)/a,\dots,(a+k-1)/k$. Other representations of $g_\alpha(x)$ and simpler forms for special values of $\alpha$ have been given by \citet{pen10:lev}.

The numerically obtained distribution is shown in Fig.~\ref{ab:phi_distri} together with the histogram for the heterogeneity parameter, $\xi$, as obtained from the numerical simulations. Clearly, the analytically derived distribution $\phi(\xi)$ from Eq.~\eqref{eq:phi} shows no agreement with the numerical result. In particular, it has to be noted that, for higher rigidities, the slope of the MSD, $\alpha$, increases, which causes the distribution $\phi(\xi)$ to be even more narrow. Instead, a simple power law provides a reasonable fit but lacks theoretical justification.

\subsection{Ergodicity breaking}

In this subsection, the non-ergodic particle behavior will be quantified in terms of a so-called ergodicity breaking parameter. According to \citet{che13:ano}, a necessary condition for ergodicity can be formulated by stating that the ratio of time and ensemble averaged MSD equals unity, i.\,e., 
\be
\m{\overline{\De^2(t)}}=\m{\De x(t)^2},
\ee
which, as emphasized earlier, is the case for cosmic-ray transport. The sufficient condition states that, in the limit of large times, the ergodicity breaking parameter
\be\label{eq:EB}
\text{EB}=\lim_{t\to\infty}\left(\bigl\langle\xi^2\bigr\rangle-\m\xi^2\right)=0
\ee
should vanish in the limit of large times.

In Fig.~\ref{ab:break} it is illustrated that this is not nearly the case. In fact, for perpendicular transport the parameter EB steadily increases instead of decreasing and is of the order 10 for large simulation times. The corresponding result for the $z$ direction, in contrast, typically remains one order of magnitude smaller. This result illustrates that, for parallel cosmic-ray scattering, the ergodicity is far less broken than for perpendicular transport.

Note that, for consistency reasons, some authors prefer to call EB a heterogeneity parameter \citep[e.\,g.,][]{thi14:bre}. A connection of the ergodicity breaking parameter to the diffusivity power index $\alpha$ has been given by \citet{hey08:msq} by requiring that $0\leqslant\text{EB}\leqslant1$. There is, however, no reason that the variance of $\xi$ be limited by one, and in fact it is not.

\subsection{Non-Gaussianity}

\begin{figure}[tb]
\centering
\includegraphics[width=\figwidth]{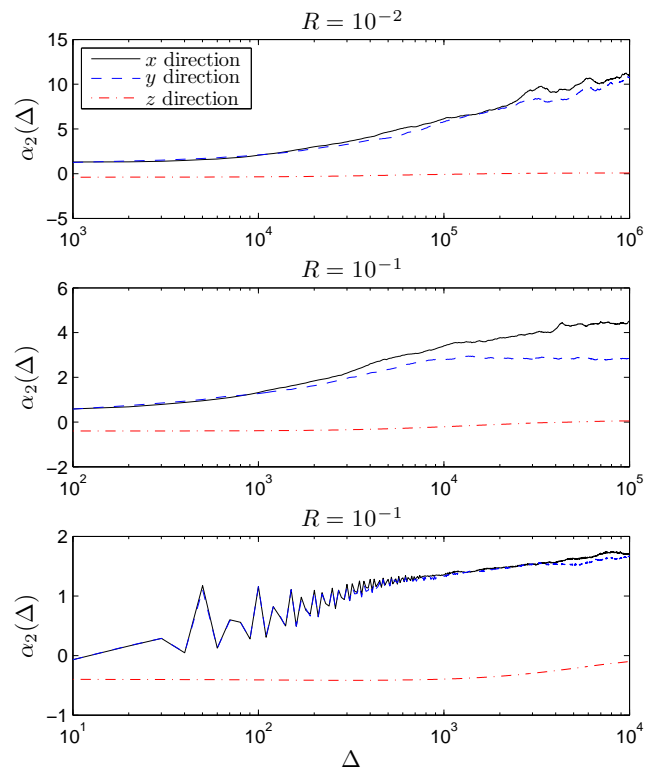}
\caption{Results for the non-Gaussianity parameter~$\alpha_2$ as defined in Eq.~\eqref{eq:gauss}. In each plot, $\alpha_2$ is shown as a function of the simulation time, $t$, for the $x$ and $y$ direction (black solid and blue dashed lines, respectively). For comparison purposes, $\alpha_2$ is also shown for the $z$ direction (red dot-dashed line).}
\label{ab:gauss}
\end{figure}

As shown in Sec.~\ref{msd:steps}, the distribution of step lengths does not follow a normal distribution. This deviation from Gaussianity can be characterized quantitatively \citep[see][]{rah64:arg,mer15:sub} through a non-Gaussianity parameter
\be\label{eq:gauss}
\alpha_2=\frac{\bigl\langle\left(\De x\se\right)^4\bigr\rangle}{\sigma\bigl\langle\left(\De x\se\right)^2\bigr\rangle^2}-1,
\ee
with $\sigma=3$ for one dimension and $\sigma=2$ for two dimensions. For a Gaussian distribution, $\alpha_2$ would be zero.

The result for the non-Gaussianity parameter $\alpha_2$ is shown in Fig.~\ref{ab:gauss}, where for comparison also the corresponding parameter for parallel transport is depicted. While the turbulent transport of charged particles in stochastic magnetic fields is essentially non-Gaussian in all directions, it turns out that parallel transport tends to become regular in the limit of large times, as shown by the approximate limit $\alpha_2\to0$ even for high rigidities.

\begin{figure}[tb]
\centering
\includegraphics[width=\figwidth]{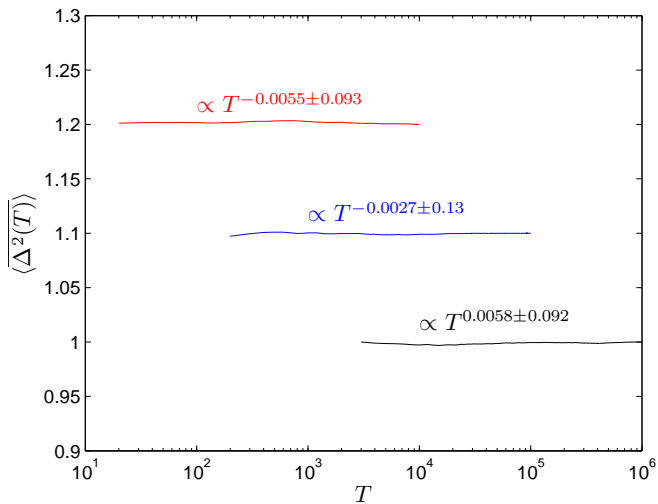}
\caption{Time-averaged MSD as a function of the measurement duration, $T$. Note that the simulation setup is such that, for lower rigidities, the longer measurement time also entails a lower time-resolution. Accordingly, the data seems to begin at later times because the first time step cannot be placed at $T=0$.}
\label{ab:tendfit}
\end{figure}

It has to be noted that parts of the results show a deviation between the $x$ and $y$ directions, which to date can be explained only by the relative importance of outliers in the calculation of the fourth moment. The uncertainties to $\alpha_2$ are thus large, while the results nevertheless indicate that the perpendicular transport is non-Gaussian.

\subsection{Aging}

If the time-averaged MSD is taken to be a function of the measurement duration, $T$, then a variation could indicate the presence of an aging process \citep{met14:erg}. Such a variation is typically observed for a broad distribution of waiting times $\tau$, since the measurement duration can only include waiting times up until the measurement time, i.\,e., $\tau\leqslant T$.

In Fig.~\ref{ab:tendfit}, the time-averaged MSD is shown as a function of $T$. For all rigidities, the results are compatible with a constant value. This result is to be expected from the fact that, as mentioned earlier, the concept of a waiting time is not applicable for continuously moving particles.

\section{Summary and Conclusion}\label{summ}

In this article, the ergodicity of cosmic-ray transport in turbulent magnetic fields has been investigated by means of numerical Monte-Carlo simulations. In particular, the transport across the mean magnetic field has been studied, which had previously been found to be subdiffusive on many occasions. In view of the controversy if, and under what circumstances, perpendicular transport behaves ``normal'', this case merits special attention.

In biology and chemistry, the characterization of particle transport processes in terms of waiting time distributions, L\'evy flights, and aging systems has a certain tradition. In high-energy astrophysics, in contrast, a diffusive---and thus ergodic---behavior is often simply assumed. Here, it has been shown that cross-field transport is, in general, weakly non-ergodic but still almost diffusive, as confirmed through several mathematical criteria. The transport is non-Gaussian, obeys the Alexander-Orbach relation, and the underlying system is non-aging as expected. The most prominent deviation from a simple ergodic process is the heterogeneity of the particle ensemble, which is characterized both by the diversity of the MSD for different turbulence realizations as well as by the heterogeneity parameter. A further interesting finding is the step length distribution, which is flat (peaked) for particles with high (low) rigidities.

For future work it may prove interesting to relate the non-ergodic behavior to the diffusion coefficients. The plethora of criteria that exist in the literature for the classification of random walks may be useful in numerical simulations as well as for observations. For instance, the validity of the Alexander-Orbach relation allows one to connect the intensity decay at a fixed point in time to the random walk through space. While beyond the scope of the present paper, it will be interesting to see if such relations hold also for more realistic turbulence models involving propagating plasma waves and transient structures.

\begin{acknowledgements}
I thank J. Pratt and the organizers of the XXXV Dynamics Days Europe conference in Exeter for inviting me to an inspiring meeting.
\end{acknowledgements}




\end{document}